\documentclass[aps,pre,onecolumn,groupedaddress,11pt]{revtex4-2}

\usepackage{amsmath}
\usepackage{graphicx}
\begin{document}
\raggedbottom

\title{Nonreciprocal McKean-Vlasov Equations: From Stationary Instabilities to Travelling Waves}


\author{Arjun R$^{1,2}$}
\author{Pratyush Prakash Patra$^{1,2}$}
\author{A. V. Anil Kumar$^{1,2}$}
\email[]{anil@niser.ac.in}
\affiliation{$^{1}$School of Physical Sciences, National Institute of Science Education and Research, Jatni, Bhubaneswar 752050, India}
\affiliation {$^{2}$Homi Bhabha National Institute, Anushakti Nagar, Mumbai, India}


\begin{abstract}
Nonreciprocal interactions, in which action-reaction symmetry is broken, provide a powerful route to collective dynamics that cannot be captured by equilibrium free-energy minimisation. Here, we introduce and analyse a two-species nonreciprocal McKean-Vlasov equation derived from an underlying system of interacting stochastic particles. Combining linear stability analysis, weakly nonlinear arguments, pseudo-spectral simulations, and Langevin particle dynamics, we show that the structure of nonreciprocity controls the onset and nature of collective order. For spatially uniform weak nonreciprocity, asymmetry shifts the critical diffusion threshold but produces only stationary instabilities, indicating that uniform imbalance alone is insufficient to generate sustained time-dependent motion. In contrast, spatially modulated nonreciprocity fundamentally enriches the dynamics: depending on its symmetry and coupling to the interaction potential, the homogeneous state can lose stability through Hopf bifurcations, giving rise to standing and travelling wave states. We identify both subcritical and supercritical Hopf transitions, relate the selected patterns to Landau saturation coefficients, and show that travelling waves can emerge even in the weak-nonreciprocity regime without explicit microscopic run-and-chase rules. Direct Langevin simulations confirm that these oscillatory and travelling states persist at the particle level and are not artefacts of the continuum mean-field description. Our results establish nonreciprocal McKean-Vlasov equations as a minimal framework for understanding how spatially structured asymmetric interactions generate self-organized motion, dynamical phase transitions, and nonequilibrium collective order.
\end{abstract}


\maketitle

\section{Introduction}

Systems of interacting stochastic particles provide a versatile and unifying framework for describing collective dynamics across a wide range of disciplines, including statistical physics, plasma physics, biological systems, and social dynamics. In many such settings, the microscopic evolution of individual constituents can be modeled as interacting diffusion processes, where each particle undergoes stochastic motion while simultaneously experiencing forces generated by all other particles in the system.  These types of models have been extensively studied in diverse contexts, including opinion dynamics \cite{castellano2009statistical}, biophysical processes \cite{bressloff2014stochastic}, cooperative phenomena in statistical mechanics \cite{dawson1983critical}, synchronization in coupled oscillators \cite{acebron2005kuramoto}, plasma dynamics \cite{rizzato2009driven}, and reaction–diffusion systems \cite{murray2003spatial}. A prototypical example of such dynamics is given by a system of $N$ interacting particles whose positions ${x^i}$, where $i \in \{1, 2, \dots, N\}$, evolve according to the stochastic differential equation
\begin{equation}
dx^i = -\frac{\kappa}{N} \sum_{j \ne i}^{N} \nabla U\left(x^i - x^j\right) dt + \sqrt{2D} dB^i(t),
\end{equation}
where $U$ is a pairwise interaction potential, $\kappa$ controls the interaction strength, $D$ is the diffusion coefficient, and $B^i(t)$ denotes independent Brownian motions. The first term represents the collective drift arising from interactions, while the second term accounts for stochastic fluctuations. In the mean-field limit $N \to \infty$, under suitable assumptions, the empirical particle density converges to a smooth one-particle distribution $\rho(x,t)$ that evolves according to the McKean-Vlasov equation (MVE)
\begin{equation}
\frac{\partial \rho}{\partial t} = D\nabla^2\rho + \kappa\nabla \cdot \bigl(\rho (\nabla U * \rho)\bigr).
\end{equation}
This equation can also be obtained as the overdamped limit of the Vlasov–Fokker–Planck equation, which motivates its nomenclature. The MVE is a nonlinear and nonlocal Fokker–Planck-type equation, and its mathematical and physical properties have been studied extensively (see, e.g., \cite{carrillo2006contractions,frank2005nonlinear,villani2021topics}). Depending on the choice of interaction potential, the McKean–Vlasov framework encompasses a variety of well-known models, including the Haken–Kelso–Bunz model in biophysics \cite{haken1985theoretical}, the noisy Kuramoto model for synchronization \cite{sakaguchi1986soluble}, and the noisy Hegselmann–Krause model for opinion dynamics \cite{pineda2013noisy}. A key feature of these equations is their ability to exhibit nontrivial collective behaviour, such as clustering, pattern formation, and the emergence of nonuniform stationary states. Importantly, the MVE possesses a gradient-flow structure, which can be written as
\begin{equation}
\frac{\partial \rho}{\partial t} = \nabla \cdot \left(\rho \nabla \frac{\delta F}{\delta \rho}\right),
\end{equation}
where $F[\rho]$ is a free-energy functional. This structure implies that the dynamics is driven by free-energy minimisation, leading generically to relaxation toward equilibrium states. Owing to this generality, the McKean–Vlasov framework has found applications across a broad spectrum of fields, ranging from stellar and plasma dynamics \cite{lynden1962stability, palmroth2018vlasov} to soft condensed matter \cite{hong2024mckean, zhang2025compound}, active matter \cite{chou2015active}, and social systems \cite{bicego2025computation, pramanik2025construction}.

In this work, we focus on a nonreciprocal extension of the MVE. Recent studies have shown that nonreciprocal interactions—where action–reaction symmetry is broken—can give rise to qualitatively new forms of collective behaviour that lie beyond the scope of traditional free-energy minimisation. These include time-dependent ordered phases, dynamical instabilities, and transitions associated with exceptional points \cite{fruchart2021non,you2020nonreciprocity,loos2023long,dadhichi2020nonmutual}. Experimental realisations further demonstrate that nonreciprocity can drive self-assembly in both biological \cite{tan2022odd} and synthetic \cite{ibele2009schooling} systems. In living systems, such asymmetric interactions often originate from information imbalances mediated by environmental factors, such as airflow, or by directional sensing mechanisms, such as vision \cite{dadhichi2020nonmutual}. Nonreciprocity has also been shown to generate emergent phenomena that are absent in equilibrium settings. For instance, polarity can arise spontaneously in mixtures of otherwise nonpolar active particles when interactions are nonreciprocal \cite{soto2014self}. More generally, systems involving multiple interacting species can exhibit complex dynamics, including periodic clustering and explosive behaviour, as observed in enzyme networks coupled through cyclic reaction pathways \cite{ouazan2023self}. These findings highlight the rich dynamical possibilities introduced by nonreciprocal couplings. On the theoretical side, a variety of minimal models have been proposed to elucidate the role of nonreciprocity in collective behaviour. Extensions of classical continuum descriptions, such as the nonreciprocal Cahn–Hilliard model, have provided insight into phase separation in asymmetric binary mixtures \cite{saha2020scalar}. At the lattice level, nonreciprocal generalisations of Ising \cite{seara2023non, avni2025nonreciprocal, arjun2026kinetic}, XY \cite{rouzaire2025nonreciprocal}, and Heisenberg \cite{bhatt2023emergent} models have revealed significant modifications to phase behaviour and critical properties. In particular, recent studies of a nonreciprocal Ising model with asymmetric interspecies couplings have identified not only conventional disordered and statically ordered phases, but also a novel time-dependent “swap” phase characterised by out-of-phase oscillations between species \cite{avni2025nonreciprocal}. Beyond mean-field theory, fluctuations and spatial structure further enrich this phenomenology: droplet excitations can destabilise static order, topological defects can disrupt oscillatory phases in low dimensions, and in higher dimensions, nonreciprocity can modify universality classes and stabilise spatiotemporal oscillations \cite{avni2025nonreciprocal}. A recent review on nonreciprocal many body physics can be found in \cite{fruchart2026nonreciprocal}.

The nonreciprocal extension of the MVE is relevant to a variety of applications where asymmetric interactions play a central role. For example, in models of opinion dynamics, the noisy Hegselmann–Krause framework \cite{pineda2013noisy} can be generalised to incorporate multiple communities with asymmetric intercommunity influence. In such settings, the underlying stochastic dynamics naturally lead, in the mean-field limit, to a nonreciprocal McKean–Vlasov description,  as shown later in section \ref{sec:NMVE}. Another example is the nonreciprocal Kuramoto model, which is a special case of MVE \cite{bertoli2024stability} and serves as a microscopic realization of a more general framework where nonreciprocal interactions can fundamentally alter collective phase behaviour, producing time-dependent chiral phases with no equilibrium counterpart. These transitions between static aligned/antialigned states and chiral states are controlled by exceptional points of the linearized dynamics and apply broadly to synchronization, flocking, and pattern formation \cite{fruchart2021non, chakraborty2025effects, ho2024nonreciprocal}.

In this work, we investigate the nonreciprocal MVE using a combination of analytical and numerical approaches, and benchmark the continuum predictions against Langevin dynamics simulations of the underlying stochastic particle system. In this manuscript, we primarily focus on the weak nonreciprocity limit, in which the interaction forces between two species remain opposite in direction but differ in magnitude \cite{fruchart2026nonreciprocal}. Through linear stability analysis, we derive the conditions under which the homogeneous state becomes unstable and distinguish between stationary and oscillatory instabilities. For the case of constant nonreciprocity, we find that the primary effect of asymmetry is to shift the critical diffusion coefficient through its dependence on the nonreciprocity parameter $\Delta$. In this regime, however, the instability remains purely stationary, with no emergence of oscillatory modes. This behaviour can be understood from the fact that weak, spatially uniform nonreciprocity does not satisfy the effective “run-and-chase” mechanism required to sustain time-dependent collective motion. In contrast, when the nonreciprocity is spatially modulated, the system exhibits a much richer range of dynamical phases. Depending on the form of the modulation, we observe nonuniform stationary states, as well as time-dependent structures such as standing and travelling waves. Notably, the emergence of travelling waves in this weakly nonreciprocal regime suggests that collective run-and-chase motion may arise even in the absence of explicit microscopic chasing dynamics, where one species is effectively driven toward the other while the other is driven away. Recent studies on systems with distance-dependent nonreciprocity include \cite{ivlev2015statistical, dinelli2023non, bartnick2016structural, fruchart2026nonreciprocal}. Our Langevin simulations further confirm that these oscillatory states are not merely mean-field artefacts but persist beyond the mean-field description.

\section{Nonreciprocal MVE}\label{sec:NMVE}
We consider a one-dimensional system composed of two species of particles, $A$ and $B$, with $N_A$ and $N_B$ particles, respectively. The particles within the same species interact through a reciprocal pair potential, while the interactions between different species are nonreciprocal. The corresponding Langevin equations are 
\begin{equation}\label{eq:Langevin}
    \dot{x}_i^\alpha = -\frac{\kappa}{N_\alpha} \sum_{k \neq i}^{N_\alpha} \frac{\partial U(x_i^\alpha - x_k^\alpha)}{\partial x_i^\alpha} +\frac{\gamma}{N_\beta}\ \sum_j^{N_\beta}  K_{\alpha \beta}(x_i^\alpha - x_j^\beta) + \sqrt{2D_\alpha} \xi_i^\alpha(t),
\end{equation}
where $\alpha,\beta\in \{A,B\}, \alpha \neq \beta$ and $K_{\alpha\beta}$ is the nonreciprocal interaction, $D_\alpha$ is the diffusion, $\kappa$ and $\gamma$ are the strengths of the interactions (taken to be one throughout the manuscript), and $\xi_i^\alpha(t)$ is the Gaussian white noise satisfying
\[
\langle \xi_i^\alpha(t) \rangle = 0, \qquad
\langle \xi_i^\alpha(t)\,\xi_j^\beta(t') \rangle = \delta_{ij}\delta_{\alpha\beta}\delta(t-t').
\]
The first term in \eqref{eq:Langevin} describes reciprocal interactions within each species, while the second term encodes asymmetric couplings between the two species. Substituting \eqref{eq:Langevin} into the Fokker-Planck equation given by
\begin{equation}  
    \frac{\partial P_N(x,t)}{\partial t} = - \sum_i \frac{\partial (\dot{x}_i^A P_N(x,t))}{\partial x_i^A} - \sum_i \frac{\partial (\dot{x}_i^B P_N(x,t))}{\partial x_i^B} + D_A \sum_i \frac{\partial^2 P_N(x,t)}{\partial (x_i^A)^2} + D_B \sum_i \frac{\partial^2 P_N(x,t)}{\partial (x_i^B)^2},
\end{equation}
where $P_N(x,t)$ is the $N (= N_A + N_B)$-particle distribution function, we get, in the mean field limit and $N_A, N_B \to \infty$ the nonreciprocal MVE
\begin{equation}\label{eq:MVE}
    \frac{\partial \rho_\alpha (x, t)}{\partial t} = D \frac{\partial ^2 \rho_\alpha (x, t)}{\partial x^2} + \frac{\partial }{\partial x} \left[ \rho_\alpha (x, t) \left(\frac{\partial U}{\partial x}* \rho_\alpha (x, t) \right)\right] - \frac{\partial }{\partial x} \left[ \rho_\alpha (x, t) \left(K_{\alpha\beta}(x)* \rho _\beta (x,t)\right)\right].
\end{equation}
Here, $\rho_\alpha(x,t)$ is the one-particle distribution function of species $\alpha$, and $*$ denotes convolution. To isolate the effects of nonreciprocity, we take the diffusion coefficients of the two species to be equal, $D_A = D_B = D$, throughout the manuscript. All dynamics are considered on a one-dimensional torus of length $2\pi$. In our model, the interspecies interaction kernel is chosen to have the form
\begin{equation}\label{eq:NonrecCoupledForce}
    K_{\alpha \beta} = -\frac{\partial \phi(x)}{\partial x} \times \begin{cases} 1 + \Delta(x),\quad \text{for  }\alpha \beta \in AB  \\ 1 - \Delta (x),\quad \text{for  }\alpha \beta \in BA\end{cases}
\end{equation}
where $\phi(x)$ is a symmetric interaction potential and $\Delta(x)$ is the nonreciprocity parameter satisfying $-1<\Delta(x)<1$. When $\Delta(x)\equiv0$, the coupling is reciprocal. For $\Delta \not\equiv 0$ or more specifically for a modulation with a nonzero even component, the cross-interactions violate pairwise action-reaction symmetry as will be shown in section \ref{sec:MVEspatialNonrec}. Such asymmetric couplings are an example of weak nonreciprocity. We first study the case of constant nonreciprocity, $\Delta(x)=\Delta$, and then extend the analysis to spatially dependent nonreciprocity. The latter allows us to investigate how spatial modulation of the asymmetry affects the stability and morphology of the emergent patterns. In addition, we also consider a variant in which nonreciprocity is coupled directly to the interaction potential rather than to the force kernel. This provides an alternative route to breaking reciprocity, but it is not identical to force-modulated nonreciprocity, since differentiating the product introduces additional terms proportional to $\phi(x)\Delta'(x)$. The intraspecies interaction potential is taken to be a $H$-unstable, even function of the form, $U(x) = \sum_ka_k\cos{(kx)}$. which is convenient for studying pattern formation and clustering because it contains multiple spatial length scales. Such multichromatic reciprocal interactions have been studied in earlier work on collective ordering and pattern selection \cite{bertoli2024stability}. For interaction potentials whose Fourier components remain non-negative (i.e., $H$-stable interactions), and in the absence of any external confinement, the free energy landscape is convex. As a result, the system admits a single global minimum corresponding to a spatially uniform density, which constitutes the unique stable stationary state \cite{bavaud1991equilibrium}.

\section{MVE with Constant nonreciprocity}
We first consider the case of constant nonreciprocity, $\Delta(x)=\Delta$, and we perform linear stability analysis (LSA) of the homogeneous steady state associated with Eq.~\eqref{eq:MVE}. Let $\rho_\alpha(x,t)= \rho_\alpha^0 + \rho_\alpha'(x,t)$ where $\rho_\alpha ^0$ is a spatially uniform reference state and $\rho_\alpha'(x,t)$ is a small perturbation. Substituting this ansatz into Eq.~\eqref{eq:MVE} and retaining only terms linear in the perturbation, we obtain
\begin{equation}\label{LinearizedMVE}
    \frac{\partial \rho_{\alpha,\beta}'}{\partial t} = D \frac{\partial^2 \rho_{\alpha,\beta}'}{\partial x^2} + \frac{\partial }{\partial x} \left[ \rho_{\alpha,\beta}^0 \left(\frac{\partial U}{\partial x}* \rho_{\alpha,\beta}' \right)\right] + \frac{\partial }{\partial x} \left[ \rho_{\alpha, \beta}^0 \left((1\pm\Delta)\frac{\partial \phi}{\partial x}* \rho_{\beta,\alpha}' \right)\right].
\end{equation}
We consider equal mean densities for the two species in order to isolate the effect of nonreciprocity, $\rho^0_{A,B} = 1/(2\pi)$. Writing Eq.~\eqref{LinearizedMVE} in terms of Fourier modes gives us
\begin{equation}\label{eq:integrals}
    \frac{\partial\rho^{\alpha,\beta}_q}{\partial t} = -Dq^2 \rho^{\alpha,\beta}_q + iq\rho^{\alpha,\beta}_0\rho^{\alpha,\beta}_q\int e^{-iqy}U'(y)dy + iq\rho^{\alpha,\beta}_0\rho^{\beta,\alpha}_q\int e^{-iqy}\phi'(y)(1\pm\Delta)dy,
\end{equation}
considering the intraspecies and interspecies interaction potentials to be of the form $U(y) = \sum_ka_k\cos{(ky)}$ and $\phi(y) = \sum_kb_k\cos{(ky)}$, respectively where $k \in \mathbf{N}$.
Utilizing this specific structure significantly generalizes the standard noisy Kuramoto model, extending its utility to diverse fields like mathematical biology \cite{faluweki2023active, painter2024biological}, liquid crystals \cite{constantin2004asymptotic,fatkullin2005critical}, and polymer dynamics \cite{frank2005nonlinear}. Furthermore, incorporating nonpositive interaction coefficients is a critical requirement, because strictly non-negative (H-stable) potentials guarantee a convex free energy where only the uniform distribution can be stable.  
Consequently, these nonpositive coefficients are necessary to break this stability and induce the continuous phase transitions required for the emergence of the multipeak stationary states we analyse.
Evaluating the integrals in Eq.\eqref{eq:integrals}, we get
\begin{equation}
\frac{\partial\rho^{\alpha,\beta}_q}{\partial t} = \left(-Dq^2 -q^2\frac{a_q}{2}\right)\rho^{\alpha,\beta}_q - (1\pm\Delta)q^2\frac{b_q}{2}\rho^{\beta, \alpha}_q.
\end{equation}
The corresponding Jacobian is
\begin{equation}\label{eq:JacobianConstNonrec}
   M_q =  \begin{bmatrix}
        -q^2\left(\frac{a_q}{2} + D \right) & -q^2(1+\Delta)\frac{b_q}{2} \\
        -q^2(1-\Delta)\frac{b_q}{2} & -q^2\left(\frac{a_q}{2} + D\right)
        \end{bmatrix}.
\end{equation}
The stability of the system is determined by the eigenvalues of $M_q$. 
For a $2\times 2$ matrix, the eigenvalues can be expressed as
\begin{equation}
\lambda_{\pm} = \frac{\mathrm{Tr}\,M}{2} 
\pm \sqrt{\left(\frac{\mathrm{Tr}\,M}{2}\right)^2 - \det M}.
\end{equation}
A stationary instability occurs when at least one eigenvalue becomes positive. This can happen in two distinct ways: either the determinant becomes negative, $\det M < 0$, or the trace becomes positive, $\mathrm{Tr}\,M > 0$.
For the Jacobian in \eqref{eq:JacobianConstNonrec}, the eigenvalues are
\begin{equation}
\lambda_\pm^q = -q^2\left(D+\frac{a_q}{2}\right) \pm q^2\frac{b_q}{2}\sqrt{1-\Delta^2}.
\end{equation}

Because $|\Delta|<1$ in the weak-nonreciprocity regime, the discriminant of the linear stability matrix remains non-negative, ensuring that the eigenvalues are purely real for all modes. As a consequence, the system cannot exhibit oscillatory (Hopf) instabilities when the nonreciprocity is spatially uniform. Instead, any loss of stability of the homogeneous state must occur through a stationary instability, characterised by the growth of a mode without temporal oscillations. In other words, while nonreciprocity modifies the growth rates of perturbations, it does not, in this case, introduce the phase lag necessary to generate complex-conjugate eigenvalues and sustained time-dependent behaviour. This distinction highlights an important limitation of weak, constant nonreciprocity: although it breaks action–reaction symmetry at the microscopic level, it does not by itself provide the dynamical feedback required for oscillatory collective motion. If $|\Delta| > 1$ (which is an example of antagonistic nonreciprocity, where species $A$ attracts $B$ but $B$ repels $A$ \cite{fruchart2026nonreciprocal}), however, the discriminant can become negative, allowing for complex-conjugate eigenvalues and thereby enabling oscillatory instabilities even in systems with spatially uniform nonreciprocity. Formally, instability can arise either when the trace of the Jacobian becomes positive or when the determinant becomes negative. The trace condition yields instability for $D < \max_q(-a_q/2)$; however, this threshold is subdominant. In practice, the determinant becomes negative at a larger value of $D$, namely when
\begin{equation}
D<\max_q\!\left[-\frac{a_q}{2}+\left|\frac{b_q}{2}\sqrt{1-\Delta^2}\right|\right],
\end{equation}
where the maximum is taken over all Fourier modes $q$. This determines the actual onset of instability.
The corresponding critical diffusion coefficient is therefore given by
\begin{equation}\label{eq:constStationaryInstability}
D_c=\max_q\left[\left|\frac{b_q}{2}\sqrt{1-\Delta^2}\right|-\frac{a_q}{2}\right].
\end{equation}
This expression shows that constant nonreciprocity shifts the threshold for pattern formation in a nonlinear manner through the factor $\sqrt{1-\Delta^2}$, even though the asymmetry parameter $\Delta$ enters the microscopic interaction linearly. Physically, increasing $|\Delta|$ effectively reduces the strength of interspecies coupling, thereby weakening the cooperative interactions responsible for collective ordering and delaying the onset of instability.

\section{MVE with spatially dependent nonreciprocity}\label{sec:MVEspatialNonrec}

We now extend the analysis to the case of spatially dependent nonreciprocity. In this setting, the asymmetry parameter $\Delta(x)$ varies in space and is conveniently represented through a Fourier expansion. The symmetry properties of $\Delta(x)$ play a crucial role in determining the structure of the linear stability matrix and, consequently, the nature of the resulting instabilities. For this reason, we analyse even and odd forms of nonreciprocity separately.

Before separating even and odd spatial modulations, it is useful to state the pair-level criterion for reciprocity. Let $x=x_A-x_B$. The cross-interaction is reciprocal if the force exerted on A by B is equal and opposite to the force exerted on B by A, namely
\[
K_{AB}(x)+K_{BA}(-x)=0.
\]
For the force-modulated kernel in Eq.~\eqref{eq:NonrecCoupledForce}, using the evenness of \(\phi\), we obtain
\[
K_{AB}(x)+K_{BA}(-x)
=
-\phi'(x)[\Delta(x)+\Delta(-x)]
=
-2\phi'(x)\Delta_e(x),
\]
where \(\Delta_e\) is the even part of \(\Delta\). Thus only the even component of the spatial modulation breaks action-reaction symmetry. A purely odd modulation of \(\Delta\) is reciprocal in this pairwise sense, even though \(K_{AB}(x)\) and \(K_{BA}(x)\) differ when compared at the same argument.

Throughout this section, we consider discrete Fourier modes labeled by the wavenumber $q$. We first focus on the case where $\Delta(x)$ is an even function of position, which we express as  $\Delta(x) =\Delta_0 \sum_{l\geq 1} c_l \cos{(lx)}$. To ensure that the system remains within the weak-nonreciprocity regime, we always impose the constraint $|\Delta_0|\sum_l |c_l| < 1$, which guarantees that the magnitude of the asymmetry remains bounded. We begin by analysing the scenario in which the nonreciprocity enters directly through the interspecies interaction kernel. Proceeding as in the previous section, in the Fourier domain, we find
\begin{align}
    \frac{\partial\rho^{\alpha,\beta}_q}{\partial t} &= -Dq^2 \rho^{\alpha,\beta}_q  - \frac{iq}{2\pi}\rho^{\alpha,\beta}_q\int e^{-iqy}\sum_kka_k\sin{(ky)}dy \\ \nonumber & - \frac{iq}{2\pi}\rho^{\beta,\alpha}_q\int e^{-iqy}\sum_kkb_k\sin{(ky)}[1\pm \Delta_0 \sum_lc_l\cos{(ly)}]dy    
\end{align}
and evaluating these integrals gives us
\begin{align}
    \frac{\partial\rho^{\alpha,\beta}_q}{\partial t} =& \left(-Dq^2  - \frac{q^2}{2} a_q\right) \rho^{\alpha,\beta}_q \\&\nonumber +\left[- \frac{q^2}{2} b_q  + \Delta_0 \frac{q}{4}\left(\mp \sum_{l=1}^{q-1} (q-l) b_{q-l} c_l \mp \sum_{l =1}^{\infty} (q +l) b_{q+l} c_l \pm \sum_{l=1}^{\infty} l b_l c_{q+l}\right) \right] \rho^{\beta,\alpha}_q
\end{align}
The corresponding Jacobian is 
\begin{equation}
   M_q =  \begin{bmatrix}
        -q^2\left(\frac{a_q}{2} + D \right) & -q^2\frac{b_q}{2} - \Delta_0 W(q) \\
        -q^2\frac{b_q}{2} + \Delta_0 W(q) & -q^2\left(\frac{a_q}{2} + D\right)
        \end{bmatrix},
\end{equation}
where $W(q) =  \frac{q}{4} \sum_{l=1}^{q-1} (q-l) b_{q-l} c_l + \frac{q}{4} \sum_{l =1}^{\infty} (q +l) b_{q+l} c_l - \frac{q}{4} \sum_{l=1}^{\infty} l b_l c_{q+l}  $ 
(where $q$ is discrete). The condition for stationary instability is found in a similar fashion as in the constant nonreciprocity case 
\begin{equation}\label{eq:StationaryInstability}
    D_c = \max_q\left[\sqrt{\frac{b_q^2}{4} - \left(\frac{\Delta_0 W(q)}{q^2}\right)^2} - \frac{a_q}{2}\right]
\end{equation}
provided that the expression under the square root is non-negative. For oscillatory instability, $\left(\frac{\text{Tr}M}{2}\right)^2 - \text{det}M < 0$ which requires
\begin{equation}\label{eq:OscCondition}
 q^4 \frac{b_q^2}{4} < [\Delta_0W(q)]^2.   
\end{equation}
Therefore, if a Fourier mode satisfies the above criterion, the corresponding eigenvalues form a complex-conjugate pair. A Hopf bifurcation of that mode occurs when the real part crosses zero, i.e. at $D_{cH}=-a_q/2$, provided $D_{cH}>0$ and provided this instability is not preempted by a stationary instability of another mode at a larger value of $D$. In contrast to the case of constant nonreciprocity, this result demonstrates that spatial modulation of the asymmetry can fundamentally alter the nature of the instability. In particular, it enables the emergence of genuine time-dependent collective states even within the weak-nonreciprocity regime, highlighting the crucial role of spatial structure in facilitating oscillatory dynamics.

Now, we consider an odd modulation, $\Delta (x) = \Delta_0 \sum_l c_l \sin{(lx)}$. In this case, the Fourier-space Jacobian becomes
\begin{equation}
   M_q =  \begin{bmatrix}
        -q^2\left(\frac{a_q}{2} + D \right) & -q^2\frac{b_q}{2} - i\Delta_0S(q) \\
        -q^2\frac{b_q}{2} + i\Delta_0S(q) & -q^2\left(\frac{a_q}{2} + D\right)
        \end{bmatrix}
\end{equation}
where $S(q) = -\frac{q}{4}\sum_k^{q-1}kb_kc_{q-k} + \frac{q}{4}\sum_{l=1}^\infty[(q+l)b_{q+l}c_l+lb_lc_{q+l}]$. In this case the eigenvalues remain real, because the discriminant contains only non-negative contributions:
\begin{equation}
\left(\frac{q^2 b_q}{2}\right)^2+[\Delta_0S(q)]^2 \ge 0.
\end{equation}
Therefore, a purely odd modulation provides a useful reciprocal control case. The corresponding linear stability matrix has conjugate off-diagonal entries and hence real eigenvalues, so it cannot generate a Hopf instability in the weak regime considered here. The only possible instability is stationary, with critical diffusion
\begin{equation}
D_c=\max_q\left[\sqrt{\frac{b_q^2}{4}+ \left(\frac{\Delta_0S(q)}{q^2}\right)^2}-\frac{a_q}{2} \right].
\end{equation}

We now consider an alternative formulation in which nonreciprocity is introduced at the level of the interaction potential itself. In this case, the interspecies interaction kernel is defined as
\begin{equation}
    K_{\alpha \beta} = -\frac{\partial}{\partial x} \times \begin{cases}  \phi(x)(1 + \Delta(x)), \quad \text{for  }\alpha \beta \in AB  \\  \phi(x)(1 - \Delta (x)), \quad \text{for  }\alpha \beta \in BA\end{cases}.
\end{equation}
Equivalently, this force contains both the prefactor-modulated term $-\phi'(x)[1\pm\Delta(x)]$ and the additional contribution $\mp \phi(x)\Delta'(x)$. Thus the condition $|\Delta(x)|<1$ bounds the asymmetry at the potential level, but does not by itself imply a pointwise weak imbalance of the resulting force. Here, we analyse both even and odd forms of the nonreciprocity function $\Delta(x)$.
For the case of even nonreciprocity, we expand
$\Delta(x) = \Delta_0\sum_k c_k \cos(kx)$.  Proceeding as in the previous analysis, we obtain the critical diffusion coefficient for the onset of stationary instability as
\begin{equation}
    D_c = \max_q\left[\sqrt{\frac{b_q^2}{4}-\left(\frac{\Delta_0B(q)}{q^2}\right)^2} - \frac{a_q}{2}\right],
\end{equation}
where
$B(q) = \frac{q^2}{4}[\sum_{l=1}^{q-1}b_{q-l}c_l + \sum_{l=1}^\infty(b_{q+l}c_l+b_lc_{q+l})]$.
In this case, oscillatory instability arises when the discriminant becomes negative, i.e.,
\begin{equation}
    \left(\frac{q^2b_q}{2}\right)^2-(\Delta_0B(q))^2<0
\end{equation}
indicating the emergence of complex-conjugate eigenvalues and hence a Hopf bifurcation.

We next consider odd modulation, taking
$\Delta(x) = \Delta_0\sum_k c_k \sin(kx)$.
Here we find similar to the previous scenario the system transforms to a reciprocal system since
\[
K_{AB}(x) + K_{BA}(-x) = -2\partial_x[\phi(x)\Delta_e(x)]
\]
and $\Delta_e(x)$ in the RHS becomes zero for purely odd-modulation.
For this case, the critical diffusion coefficient for stationary instability is given by
\begin{equation}
    D_c = \max_q\left[\sqrt{\frac{b_q^2}{4} + \left(\frac{\Delta_0F(q)}{q^2}\right)^2} - \frac{a_q}{2}\right],
\end{equation}
where
$F(q) = \frac{q^2}{4}[-\sum_{m=1}^{q-1}b_mc_{q-m} - \sum_{m=1}^\infty(b_mc_{m+q}-b_{m+q}c_m)]$.
In contrast to the even case, oscillatory instability is not possible here, since the condition
\begin{equation}
    \left(\frac{q^2b_q}{2}\right)^2+[\Delta_0F(q)]^2 < 0 
\end{equation}
cannot be satisfied for real-valued coefficients. Therefore, we conclude that the odd modulation--whether introduced at the level of the interaction potential or the force--does not give rise to oscillatory solutions in the weak-nonreciprocity limit, and only stationary instabilities can occur.

\subsection{Normal-form interpretation}

While LSA predicts the critical diffusion $D_{cH}$ at which the homogeneous state loses stability via a Hopf bifurcation, a normal-form description near the instability provides a useful interpretation of the numerical results. Near $D_{cH}$, the local density fluctuations of our nonreciprocal mixture can be represented by slowly varying left- and right-travelling wave amplitudes, $A_L(t)$ and $A_R(t)$, corresponding to the critical wavenumber $q$. Their dynamics are governed by the coupled Stuart-Landau equations \cite{stuart1960non,stewartson1971non}:
\begin{align}
    \frac{dA_R}{dt} &= (\sigma + i \omega )A_R - g_s |A_R|^2 A_R - g_c |A_L|^2 A_R, \\
    \frac{dA_L}{dt} &= (\sigma + i \omega ) A_L - g_s |A_L|^2 A_L - g_c |A_R|^2 A_L,
\end{align}
where $\sigma \propto (D_{cH} - D)$ is the linear growth rate, and the complex Landau coefficients $g_s$ (self-saturation) and $g_c$ (cross-saturation) determine the nonlinear saturation of the modes.

The relative magnitudes of the real parts, $\text{Re}[g_s]$ and $\text{Re}[g_c]$, dictate the selection between the travelling and standing waves observed in our numerical simulations. A travelling-wave state, characterized by polar order where one direction dominates (e.g., $|A_R| > 0, A_L = 0$), emerges when cross-saturation heavily penalizes coexistence ($\text{Re}[g_c] > \text{Re}[g_s]$). This bifurcation is supercritical if $\text{Re}[g_s] > 0$, explaining the continuous emergence of the travelling waves observed for the $q=2$ mode in our system. 

Conversely, a standing-wave state preserves left-right symmetry ($|A_R| = |A_L| > 0$) and occurs when self-saturation dominates ($\text{Re}[g_s] > \text{Re}[g_c]$). The criticality of the standing wave depends on the combined saturation coefficient: it is supercritical if $\text{Re}[g_s + g_c] > 0$, but subcritical if $\text{Re}[g_s + g_c] < 0$. In the latter case, the cubic terms fail to arrest exponential growth, forcing the system to undergo a discontinuous jump to a finite-amplitude state. This framework directly explains the subcritical Hopf bifurcation and the resulting standing-wave oscillatory state observed for the $q=1$ mode, highlighting how spatially dependent nonreciprocity can selectively drive distinct macroscopic phases.

By systematically deriving $\text{Re}[g_s]$ and $\text{Re}[g_c]$ from the microscopic interaction potentials $U(x)$, $\phi(x)$, and the nonreciprocity parameter $\Delta(x)$, one can rigorously map the phase boundaries of the nonreciprocal MVEs.

\section{Numerical simulations}
To numerically solve the coupled nonreciprocal MVEs Eq.~\eqref{eq:MVE}, we employ a pseudo-spectral numerical method \cite{trefethen2000spectral, boyd2001chebyshev}. The one-dimensional spatial domain is discretised into a uniform grid, and the density fields of both interacting species are transformed into Fourier space to efficiently evaluate spatial derivatives and non-local convolution integrals using Fast Fourier Transforms (FFTs). For time integration, the system utilizes a semi-implicit Euler scheme. Specifically, the nonlinear drift components---driven by the gradients of the reciprocal and nonreciprocal interaction potentials---are evaluated explicitly, while the linear diffusion operator is treated implicitly to maintain numerical stability over extended integration times. This approach allows us to accurately track the system's intermediate and long-time dynamics, facilitating the study of emergent collective behaviours and non-uniform stationary states as the diffusion coefficient $D$ is varied \cite{chen1998applications}.

We begin by considering the case of constant nonreciprocity and use it as a benchmark to validate the predictions obtained from LSA. In particular, we compare the numerically observed transition from a homogeneous to a patterned state with the analytically predicted critical diffusion coefficient.
\begin{figure}
    \includegraphics[width=0.95\linewidth]{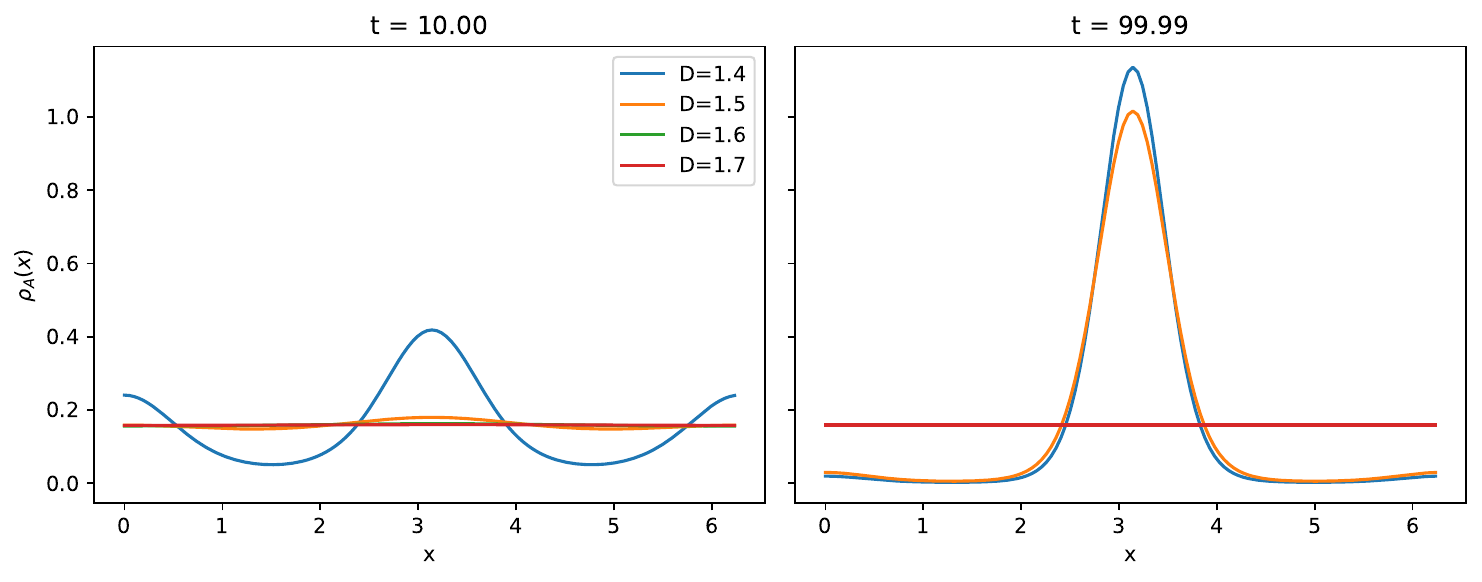}
    \caption{Time evolution of the density profile $\rho_A(x)$ at $t=10.00$ and $t=99.99$ for a system with constant nonreciprocity. At the critical diffusion point $D_c = 1.508$, the uniform state loses stability. For values $D > D_c$, the system relaxes back to a homogeneous steady state, whereas for $D < D_c$, it evolves into a non-uniform stationary state. Parameters: $U(x) = -2.0 \cos x - \cos 2x$, $\phi(x) = -\cos x - 2.2 \cos 2x$, and $\Delta = 0.4$.\label{fig:constNonrec1}}
\end{figure}
\begin{figure}
    \includegraphics[width=0.95\linewidth]{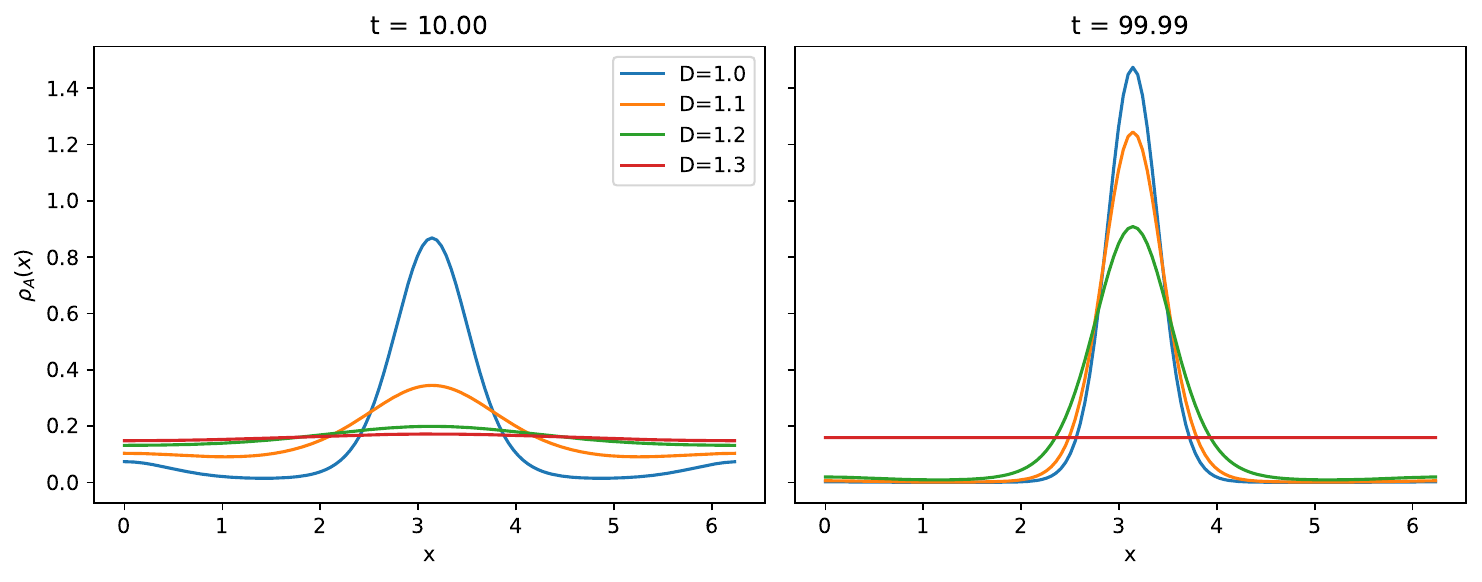}
    \caption{Time evolution of the density profile $\rho_A(x)$ at $t=10.00$ and $t=99.99$ with a higher constant nonreciprocity. Similar to Figure 1, the system relaxes to a homogeneous steady state for $D > D_c = 1.218$ and forms a non-uniform stationary state for $D < D_c$. Parameters: $U(x) = -2.0 \cos x - \cos 2x$, $\phi(x) = -\cos x - 2.2 \cos 2x$, and $\Delta = 0.9$.\label{fig:constNonrec2}}
\end{figure}
To probe both the intermediate- and long-time dynamics, we consider representative choices of the interaction potentials $U(x)$ and $\phi(x)$. In this section, we focus on $H$-unstable potentials, for which the homogeneous state is susceptible to instability and can give rise to structured steady states. A typical example is given by
\[
U(x)=-2.0\cos{x}-1.0\cos{2x}, \qquad
\phi(x)=-1.0\cos{x}-2.2\cos{2x}.
\]
For systems with constant nonreciprocity, the uniform steady state is taken as $\rho_0^{A,B} = 1/(2\pi)$. To examine its stability, we perturb this state by introducing a small sinusoidal modulation of amplitude $0.005$ in both species.
Due to the translational invariance of the governing equation, the eventual location of density peaks in the non-uniform state depends on the initial perturbation. In the simulations presented here, the phase of the perturbation is chosen such that the resulting structure is centered at $x = \pi$.
For $\Delta = 0.4$, from the analytical result in Eq.~\eqref{eq:constStationaryInstability}, we find that the $q=2$ mode becomes unstable via a stationary bifurcation at the critical diffusion coefficient $D_c = 1.508$. The numerical results are in excellent agreement with this prediction. As illustrated in Figs.~\ref{fig:constNonrec1}, for $D > D_c$ the system relaxes back to the homogeneous steady state, indicating stability of the uniform configuration. At $D = D_c$, the system undergoes a loss of stability, and for $D < D_c$ the dynamics evolve toward a non-uniform stationary state characterised by spatial density modulation. We perform another simulation with the same potentials for $\Delta = 0.9$, however, here we find that $q=1$ mode undergoes stationary instability where the critical diffusion, $D_c = 1.218$. These are in complete agreement with LSA results as is evident in Fig.~\ref{fig:constNonrec2}. The close agreement between the numerical and analytical values of the critical diffusion coefficient provides strong validation of the LSA and confirms the consistency of the continuum description.

\begin{figure}
    \includegraphics[width=0.95\linewidth]{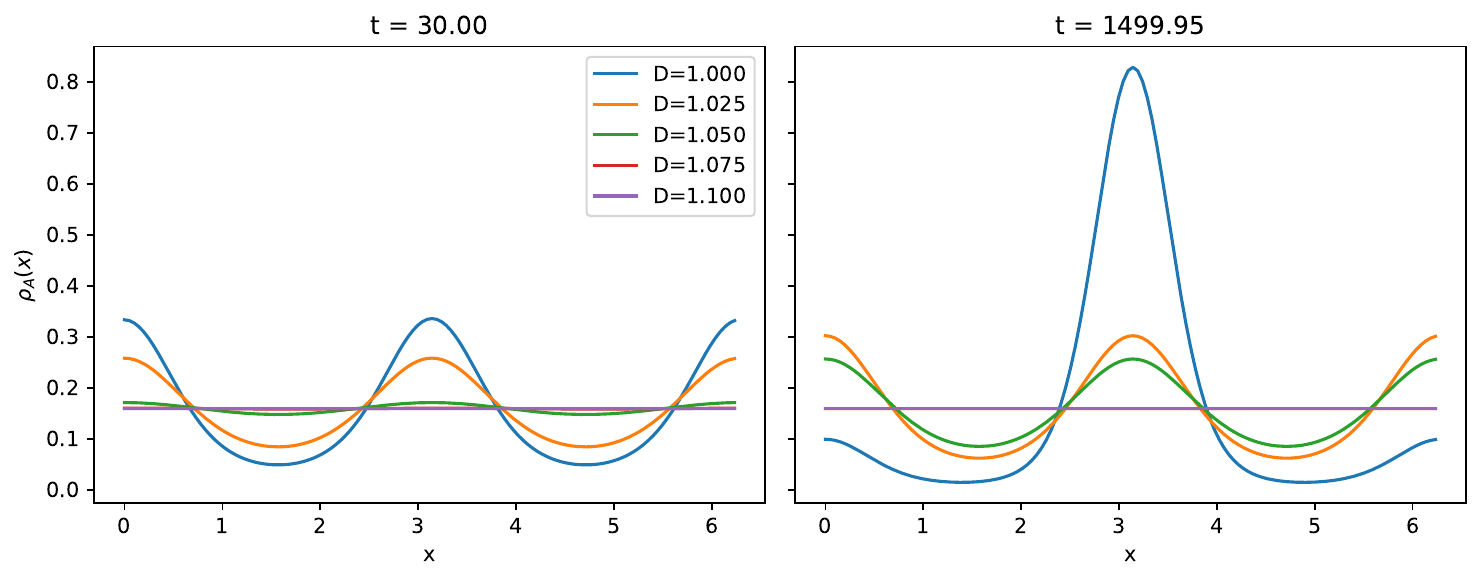}
    \caption{Pseudo-spectral simulation results illustrating the time evolution of the density profile $\rho_A(x)$ under spatially dependent nonreciprocity where only stationary instability is present. As the diffusion coefficient $D$ is decreased, the $q=2$ mode is the first to become unstable at the critical value $D_c = 1.073$. The system decays to a homogeneous steady state for $D > D_c$ and relaxes to a non-uniform stationary profile for $D < D_c$. Parameters: $U(x) = -1.0 \cos x - 2.0 \cos 2x$, $\phi(x) = -0.6 \cos x - 0.2 \cos 2x$, and $\Delta = 0.9 \cos x$. \label{fig:NonrecSpatialStationary}}
\end{figure}

We next turn to the case of spatially dependent nonreciprocity and, for simplicity, consider a cosine modulation of the form $\Delta(x) = 0.9\cos x$. As a representative scenario, we focus on the case in which the nonreciprocity enters through the interaction force, allowing us to directly examine its impact on the stability of the system. We begin by selecting interaction potentials for which the LSA predicts the absence of oscillatory instabilities across all Fourier modes. A representative example is given by
\[
    U(x)=-1.0\cos{x}-2.0\cos{2x}, \qquad
\phi(x)=-0.6\cos{x}-0.2\cos{2x}.
\]
For this choice of parameters, the coefficients $b_1$ and $b_2$ do not satisfy the oscillatory-instability condition \eqref{eq:OscCondition} for any wavenumber $q$. As a result, the system exhibits only stationary instability, consistent with the analytical predictions. The critical diffusion coefficient is therefore determined by the stationary-instability condition given in \eqref{eq:StationaryInstability}. Evaluating this expression shows that the instability threshold is governed by the mode $q = 2$, for which the right-hand side of \eqref{eq:StationaryInstability} attains its maximum. This yields a critical value $D_c = 1.073$. Consequently, as the diffusion coefficient $D$ is reduced, the $q = 2$ mode is the first to become unstable, leading to the emergence of spatially structured stationary states.
Fig.~\ref{fig:NonrecSpatialStationary} shows the pseudo-spectral simulation results. For $D>D_c$, the system decays to the homogeneous steady state, whereas for $D<D_c$ it relaxes to a non-uniform stationary profile.
\begin{figure}
    \includegraphics[width=0.95\linewidth]{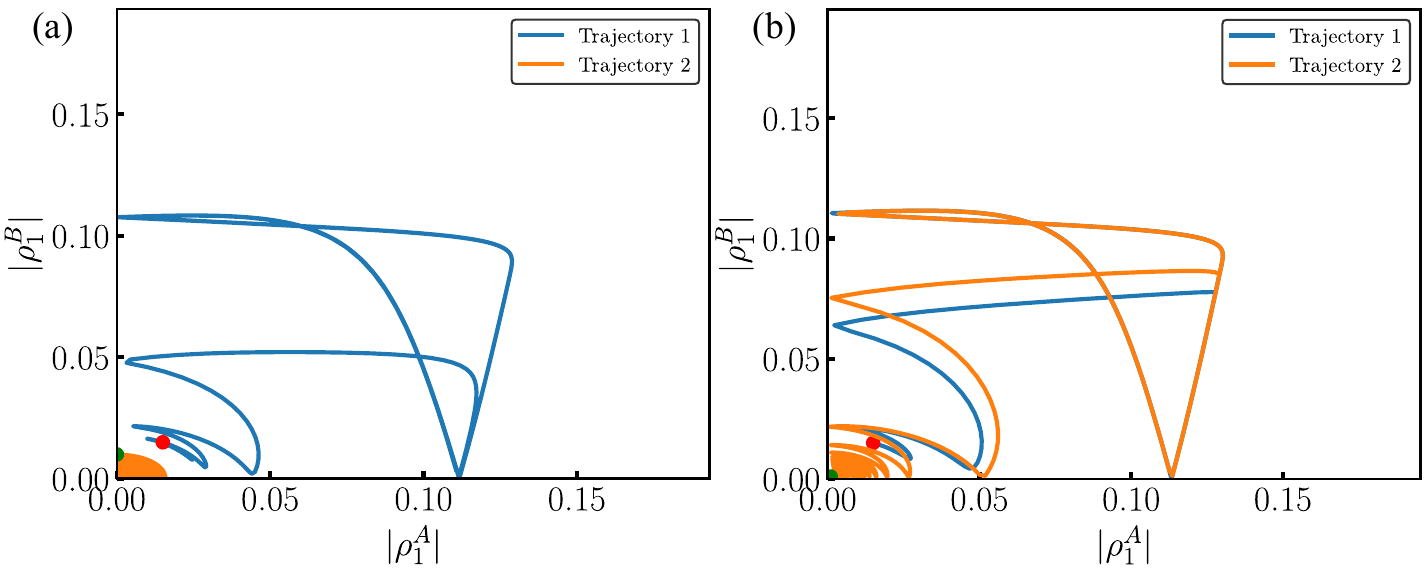}
    \caption{Phase space trajectories illustrating a subcritical Hopf bifurcation for the $q=1$ mode. Red and green markers indicate the initial points of the trajectories 1 and 2, respectively. (a) At a diffusion coefficient of $D=1.51$ ($D > D_{cH}$), trajectories exhibit a coexistence of stable and unstable limit cycles, characteristic of a saddle-node bifurcation of limit cycles. Initial conditions with a radius close to zero (Trajectory 2) decay toward the origin, whereas sufficiently large perturbations (Trajectory 1) converge to a limit cycle centered at the origin. (b) At $D=1.499$ ($D < D_{cH}$), the discontinuous transition is evident as all trajectories converge to the outer stable limit cycle. Parameters: $U(x)=-3.0 \cos x - 2.0 \cos 2x$, $\phi(x)=-0.2 \cos x - 0.5 \cos 2x$, $\Delta=0.9 \cos x$, and $D_{cH}=1.5$. \label{fig:subModrho}}
\end{figure}
\begin{figure}
    \includegraphics[width=0.95\linewidth]{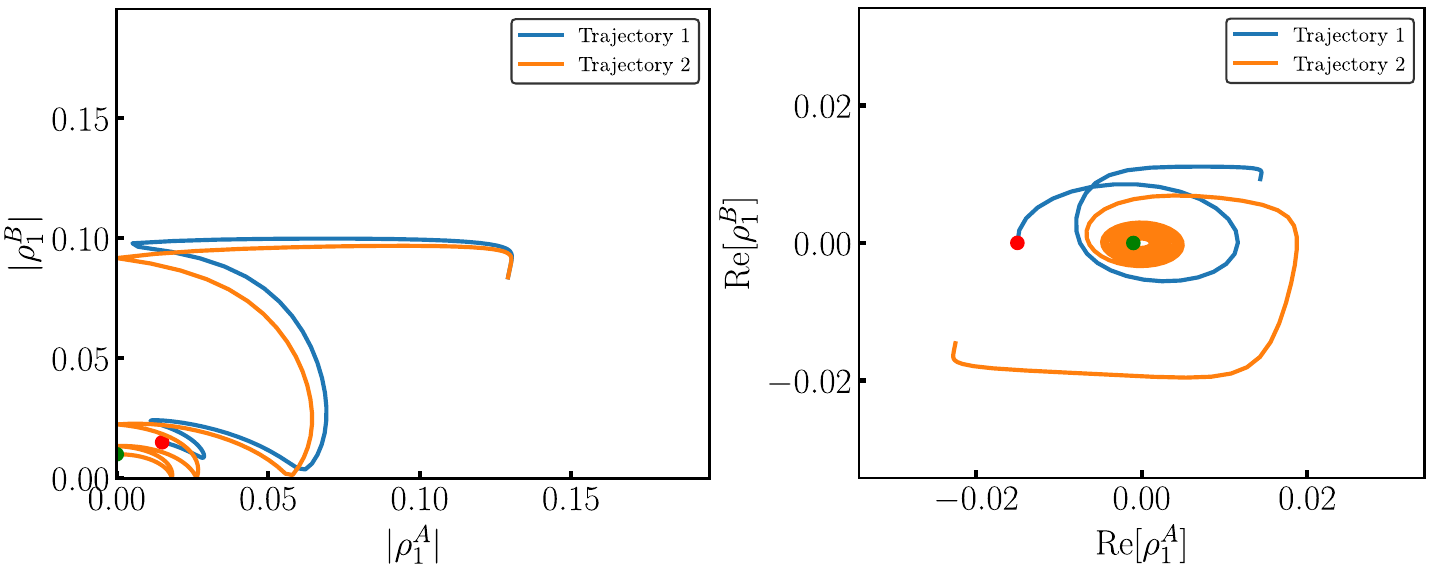}
    \caption{Phase space trajectories demonstrating a saddle-node on invariant circle bifurcation. (a) Upon further decreasing the diffusion coefficient to $D=1.494$, both trajectories asymptotically approach a constant radial amplitude. (b) The phase portrait in the $\text{Re}[\rho_1^A]$–$\text{Re}[\rho_1^B]$ plane confirms that the complex amplitudes $\rho_1^\alpha$ asymptotically approach a stable fixed point within their four-dimensional phase space. Parameters: $U(x)=-3.0 \cos x - 2.0 \cos 2x$, $\phi(x)=-0.2 \cos x - 0.5 \cos 2x$, $\Delta=0.9 \cos x$, and $D_{cH}=1.5$.\label{fig:subSnic}}
\end{figure}

We now consider interaction potentials for which oscillatory instabilities are possible. For simplicity, we take 
\begin{equation}\label{eq:PotentialForm}
U(x)=a_1\cos{x}+a_2\cos{2x}, \qquad
\phi(x)=b_1\cos{x}+b_2\cos{2x}, \qquad
\Delta(x)=c_1\cos{x}.
\end{equation}
Using the LSA criterion for oscillatory instability, we find that such an instability can occur only if any Fourier mode $q$ satisfies
\begin{equation}
    \left(\frac{q^2 b_q}{2}\right)^2 < c_1^2 \left( \frac{q (q+1)}{4} b_{q+1} + \frac{q (q-1)}{4} b_{q-1} \right)^2.
\end{equation}

As a representative example, we consider the parameter set
\begin{equation}\label{eq:q1Osc}
    a_1=-3.0, \qquad a_2=-2.0, \qquad b_1=-0.2, \qquad b_2=-0.5, \qquad c_1=0.9.
\end{equation}
For these values, the LSA predicts qualitatively distinct behaviour for different Fourier modes. In particular, the $q=2$ mode does not satisfy the oscillatory-instability condition and therefore undergoes a purely stationary instability. In contrast, the $q=1$ mode satisfies the criterion for oscillatory instability and consequently exhibits a Hopf bifurcation at the critical diffusion coefficient $D_{cH} = -a_1/2$. Since LSA provides only local information about the stability of the homogeneous state, we complement it with pseudo-spectral partial differential equation (PDE) simulations to characterise the nonlinear dynamics and determine the nature of the bifurcation. To this end, we focus on the $q=1$ Fourier component of the density fields, denoted by $\rho_{q=1}^\alpha$ with $\alpha \in \{A,B\}$. Although this mode is dynamically coupled to higher harmonics, its evolution captures the essential features of the bifurcation. Because $\rho_{q=1}^A$ and $\rho_{q=1}^B$ are complex-valued, the dynamics unfolds in an effective four-dimensional phase space. For diffusion coefficients above the Hopf threshold ($D = 1.51 > D_{cH}$), we observe from Fig.~\ref{fig:subModrho}(a) that small-amplitude perturbations decay toward the homogeneous state (Trajectory 2), while sufficiently large initial conditions evolve toward a finite-amplitude limit cycle (Trajectory 1). This coexistence of a stable fixed point with a stable limit cycle, separated by an unstable cycle, is indicative of a saddle-node bifurcation of limit cycles. As $D$ approaches $D_{cH}$ from above, the radius of the unstable limit cycle decreases, eventually merging with the fixed point at the bifurcation threshold. Taken together with the linear stability results, this behaviour is consistent with a subcritical Hopf bifurcation at $D = D_{cH}$. At $D = 1.499$, as shown in Fig.~\ref{fig:subModrho}(b), all trajectories converge to the outer stable limit cycle, reflecting the discontinuous nature of the transition. This is further corroborated by the behaviour of the order parameter as a function of $D$. Upon decreasing $D$ further, the system undergoes a saddle-node on invariant circle (SNIC) like bifurcation. As shown in Fig.~\ref{fig:subSnic}(a), for $D=1.494$, both trajectories asymptotically approach a constant radial amplitude. In conjunction with Fig.~\ref{fig:subSnic}(b), this confirms that the system ultimately converges to a stable fixed point within the four-dimensional phase space of the complex amplitudes $\rho_1^\alpha$.
This subcritical Hopf bifurcation in the \(q=1\) mode arises from the combined effect of self- and cross-saturation, for which \(\mathrm{Re}[g_s] + \mathrm{Re}[g_c] < 0\). Because the mode corresponds to a standing wave, this negative contribution implies that small perturbations grow away from the homogeneous steady state at the Hopf threshold.

\begin{figure}
    \includegraphics[width=0.95\linewidth]{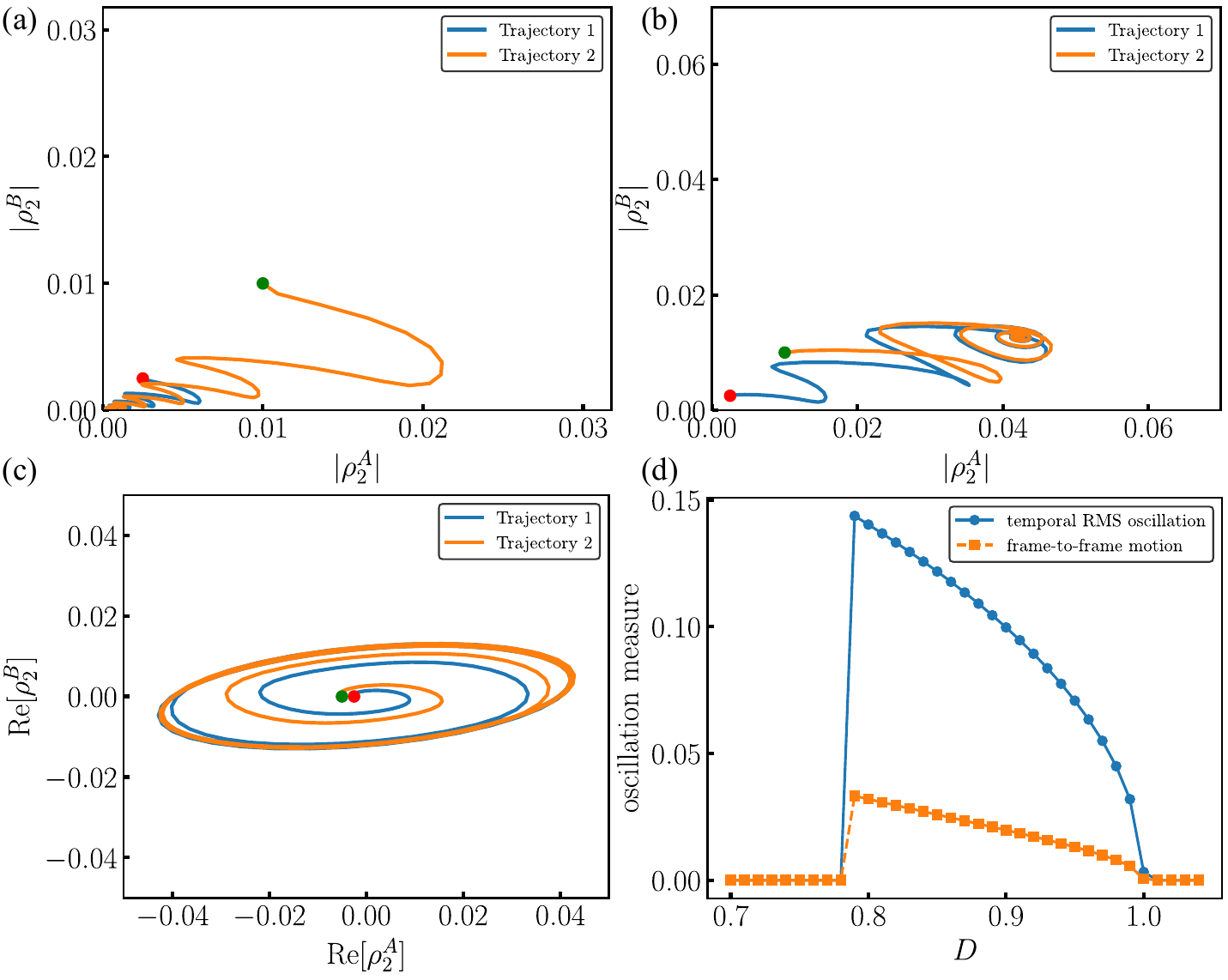}
    \caption{Dynamics of a supercritical Hopf bifurcation and sustained oscillations for the $q=2$ mode. (a) For $D=1.01$ ($D > D_{cH}$), trajectories in the four-dimensional phase space decay to zero. (b) For $D=0.98$ ($D < D_{cH}$), trajectories approach a small region around a non-zero radius. (c) A parametric plot of $\text{Re}[\rho_2^{A}]$ versus $\text{Re}[\rho_2^{B}]$ illustrates the birth of a finite-amplitude limit cycle for $D < D_{cH}$, consistent with a supercritical Hopf bifurcation. (d) The oscillation measure plotted as a function of $D$ shows the continuous emergence of sustained oscillations at the critical value $D_{cH}=1.0$. The amplitude of the limit cycle increases progressively as $D$ decreases, until the system undergoes a SNIC-like bifurcation at lower $D$ values, beyond which oscillations cease. Parameters: $U(x)=-0.4 \cos x - 2.0 \cos 2x$, $\phi(x)=-0.8 \cos x - 0.15 \cos2 x$, $\Delta=0.9 \cos x$, and $D_{cH}=1.0$.\label{fig:supModrho}}
\end{figure}

We next consider a parameter regime in which the $q=2$ mode satisfies the oscillatory-instability condition. Specifically, we choose
\[
a_1=-0.4, \qquad a_2=-2.0, \qquad b_1=-0.8, \qquad b_2=-0.15, \qquad c_1=0.9.
\]
For this set of parameters, the $q=2$ mode fulfils the oscillatory criterion, and a Hopf bifurcation is predicted to occur at the critical diffusion coefficient $D_{cH} = -a_2/2 = 1.0$.
To determine the nature of this bifurcation, we analyse the nonlinear dynamics using pseudo-spectral PDE simulations. For $D = 1.01$, with initial conditions taken far from the homogeneous state, the trajectories in the four-dimensional phase space decay toward the origin, indicating stability of the uniform solution (see Fig.~\ref{fig:supModrho}(a)). In contrast, for $D = 0.98$, the trajectories evolve toward a finite-radius attractor, as shown in Fig.~\ref{fig:supModrho}(b). This behaviour is further illustrated in Fig.~\ref{fig:supModrho}(c), where a phase-space projection of $\mathrm{Re}[\rho_2^B]$ versus $\mathrm{Re}[\rho^A_2]$ reveals the emergence of a closed orbit corresponding to a finite-amplitude limit cycle.
The continuous onset of this limit cycle for $D < D_{cH}$ is characteristic of a supercritical Hopf bifurcation associated with the $q=2$ mode. Additionally, the time evolution obtained from the PDE simulations shows that the system initially forms a transient standing-wave pattern, which subsequently evolves into a stable travelling-wave state. This crossover highlights the rich nonlinear dynamics that can arise even in the vicinity of a supercritical bifurcation. This supercritical Hopf bifurcation in the $q=2$ mode with the travelling-wave nature of the instability implies that self-saturation coefficient satisfies $\mathrm{Re}[g_s] > 0$. Consequently, small perturbations decay at the Hopf threshold, and as $D$ decreases, a finite-amplitude travelling-wave state emerges continuously from the homogeneous steady state.

\begin{figure}
    \includegraphics[width=0.95\linewidth]{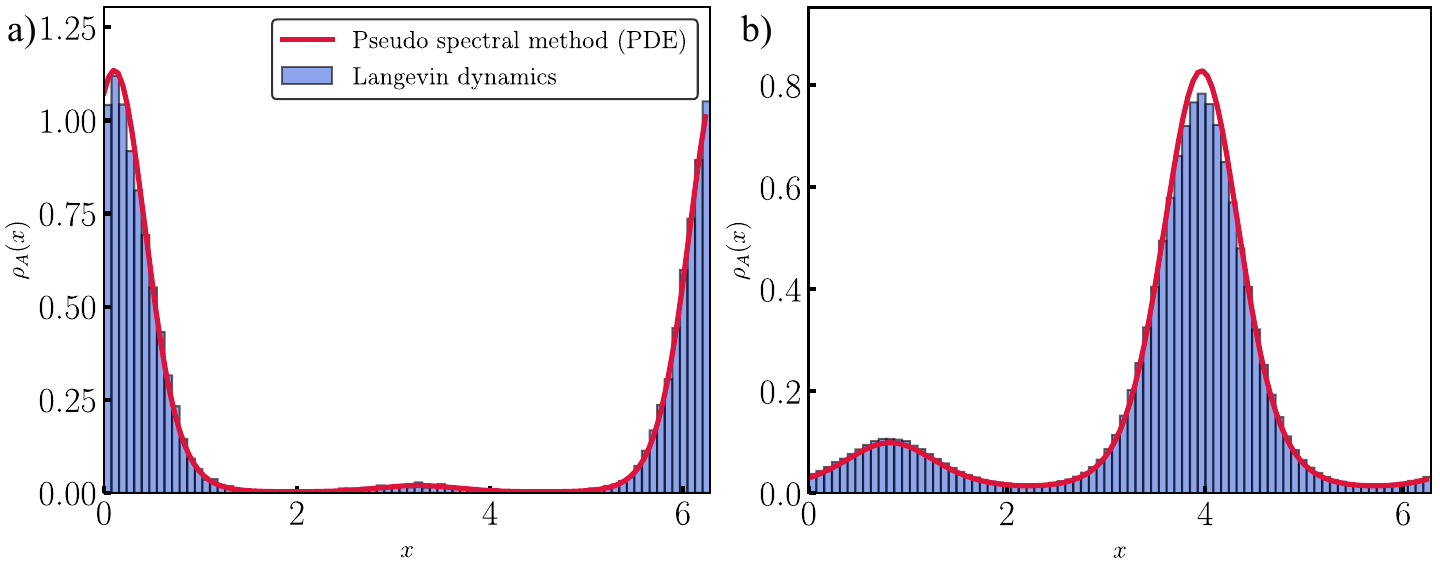}
    \caption{(a) Comparison of the density profile $\rho_A(x)$ obtained from the pseudo-spectral method and Langevin dynamics simulations for constant nonreciprocity. The Langevin stochastic equations show good agreement with the continuum mean-field PDE results. Parameters: $U(x) = -2.0 \cos x - 1.0 \cos 2x$, $\phi(x) = -1.0 \cos x - 2.2 \cos 2x$, $\Delta = 0.4$, $D_c=1.508$, $D=1.4$, and $N_A = N_B = 4000$. (b) Comparison of the density profile $\rho_A(x)$ obtained from PDE and Langevin dynamics simulations for spatially dependent nonreciprocity. For interaction potentials that do not satisfy the oscillatory-instability condition, the system exhibits only stationary instability, with the particle-level simulations demonstrating excellent agreement with the PDE simulations. Parameters: $U(x) = -1.0 \cos x - 2.0 \cos 2x$, $\phi(x) = -0.6 \cos x - 0.2 \cos 2x$, $\Delta = 0.9 \cos x$, $D_c=1.073$, $D=1.0$, and $N_A = N_B = 4000$. \label{fig:LangevinNonrec}}
\end{figure}

To quantify the oscillatory dynamics, we monitored the density fields $\rho_A(x,t)$ and $\rho_B(x,t)$ after discarding an initial transient and retained snapshots at regular intervals in the late-time regime. The oscillation amplitude was measured from the root-mean-square of the temporal fluctuations about the time-averaged profile,
\[
\delta \rho_\alpha(x,t)=\rho_\alpha(x,t)-\langle \rho_\alpha(x,t)\rangle_t, \qquad \alpha \in \{A,B\},
\]
and
\[
A_{\mathrm{osc}}=
\left[
\frac{1}{2}\left(
\langle \delta\rho_A^2\rangle_{x,t}
+
\langle \delta\rho_B^2\rangle_{x,t}
\right)
\right]^{1/2},
\]
where $\langle \cdot \rangle_{x,t}$ denotes averaging over space and over the post-transient sampled times. In addition, we computed a direct measure of dynamical motion from successive sampled configurations,
\[
M=
\left[
\frac{1}{2}\left(
\left\langle \big(\rho_A^{n+1}-\rho_A^{n}\big)^2\right\rangle_x
+
\left\langle \big(\rho_B^{n+1}-\rho_B^{n}\big)^2\right\rangle_x
\right)
\right]^{1/2},
\]
which serves as a complementary indicator of sustained temporal oscillations in the steady state. Figure~\ref{fig:supModrho}(d) presents the oscillation measure as a function of $D$. A clear onset of sustained oscillations is observed at the critical value $D_{cH} = 1.0$, consistent with a supercritical Hopf bifurcation, where a limit cycle of vanishingly small amplitude emerges continuously from the homogeneous steady state. Upon further decreasing $D$, the amplitude of the limit cycle increases progressively, indicating the growth of nonlinear oscillatory dynamics. At lower values of $D$, the system undergoes a SNIC-like bifurcation, beyond which the oscillations cease, and the dynamics converge to a nonuniform steady state.

\subsection{Langevin dynamics simulations}

Since the MVE is a mean-field description, the PDE simulations are strictly valid in the thermodynamic mean-field limit. To verify whether the predicted states persist in the underlying particle system, we also perform Langevin dynamics simulations using the corresponding stochastic differential equations \eqref{eq:Langevin}. The system consists of two interacting particle species, A and B, confined to a one-dimensional periodic domain of length $2\pi$. Initial particle positions are generated via rejection sampling from a weakly perturbed homogeneous density profile. The time evolution of the particle trajectories, governed by the corresponding stochastic differential equations, is numerically integrated using the Euler-Maruyama scheme \cite{maruyama1955continuous, higham2001algorithmic}. At each discrete time step (time step for all the Langevin dynamics simulation is taken to be $\Delta t = 0.005$), the position of every particle is updated by evaluating the deterministic drift---which incorporates both the reciprocal intraspecies forces and the nonreciprocal interspecies couplings---alongside a stochastic displacement driven by Gaussian white noise scaled by the diffusion coefficient $D$. Periodic boundary conditions are strictly enforced after each step, allowing us to explicitly track the emergence of dynamic collective behaviours and spatial patterns over the total simulation time. 

For the case of constant nonreciprocity, we find excellent agreement between the Langevin simulations and the PDE results. A representative comparison is shown in Fig.~\ref{fig:LangevinNonrec}(a) (parameter values for the simulation is provided in the caption), where both approaches exhibit the same transition from a homogeneous to a patterned stationary state at the predicted critical diffusion coefficient. We then extend this comparison to the case of spatially dependent nonreciprocity, focusing first on parameter regimes where the oscillatory-instability condition \eqref{eq:OscCondition} is not satisfied for any Fourier mode. In this regime, the system undergoes only stationary instability, with the threshold determined by Eq.~\eqref{eq:StationaryInstability}. As illustrated in Fig.~\ref{fig:LangevinNonrec}(b), simulated for the parameter values given in the caption, the particle-based simulations again show quantitative agreement with the continuum predictions.

We next consider a parameter regime in which the $q=1$ mode satisfies the oscillatory-instability condition. The interaction potential is chosen according to \eqref{eq:PotentialForm}, with parameters specified in \eqref{eq:q1Osc}. For the Langevin simulation we take $N_A = N_B = 8000$. Movie 1 and Movie 2 in the ancillary materials (as anc/Movie1 and anc/Movie2) provide a direct comparison between the PDE and Langevin simulations, respectively, for this case. For diffusion coefficients exceeding the Hopf threshold ($D = 1.51 > D_{cH} = 1.5$), the system exhibits standing-wave oscillations, demonstrating strong agreement between the discrete particle-level dynamics and the continuum PDE simulations. This consistency confirms that the system undergoes a subcritical Hopf bifurcation at $D = D_{cH}$. This implies that the transition to oscillatory state is discontinuous, as evidenced by the sudden onset of finite-amplitude oscillations, indicating that it corresponds to a first-order–like dynamical phase transition. When $D$ is reduced further a SNIC-like bifurcation takes place in this particular example.

Finally, we consider a parameter regime in which only the $q=2$ mode satisfies the oscillatory-instability condition. The interaction potentials are chosen in the form \eqref{eq:PotentialForm} for which the parameter values are taken as
\[
a_1=-0.4, \qquad a_2=-2.0, \qquad b_1=-0.8, \qquad b_2=-0.005, \qquad c_1=0.9.
\]
For this choice, the LSA predicts a Hopf bifurcation of the $q=2$ mode at the critical diffusion coefficient $D_{cH} = 1.0$.
The pseudo-spectral PDE simulations reveal that, for $D = 0.8 < D_{cH}$, the system develops a travelling-wave state, as illustrated in Movie 3 (anc/Movie3). This result is particularly noteworthy, as it demonstrates that coherent travelling structures can emerge even in the weak-nonreciprocity regime, without the need for explicit run-and-chase dynamics typically associated with self-propelled systems. To verify that this behaviour is not an artefact of the mean-field approximation, we perform corresponding Langevin dynamics simulations with $N_A = N_B = 8000$ and at $D = 0.8$. As shown in Movie 4 (anc/Movie4), the travelling-wave state persists at the particle level, confirming the robustness of this phenomenon. As the diffusion coefficient is decreased below $D_{cH}$, we observe that the amplitude of the oscillations grows continuously from zero. This behaviour is consistent with the normal form of a supercritical Hopf bifurcation, in which a stable limit cycle emerges smoothly from the fixed point and its amplitude increases monotonically with the distance from the bifurcation threshold.
\begin{figure}
    \includegraphics[width=0.95\linewidth]{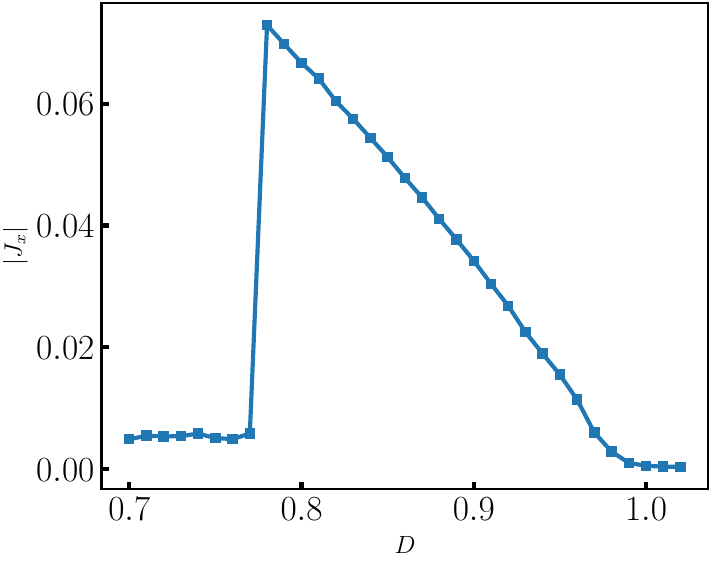}
    \caption{Dependence of the flux order parameter $|J_x|$ on the diffusion coefficient $D$ in the travelling-wave regime. For the parameter set $U(x)=-0.4\cos x-2.0\cos 2x$, $\phi(x)=-0.8\cos x-0.005\cos 2x$, and $\Delta(x)=0.9\cos x$, the flux vanishes above the Hopf threshold $D_{cH}=1.0$ and increases continuously as $D$ is lowered below $D_{cH}$, indicating the onset of directed motion and travelling-wave order. A SNIC-like bifurcation occurs as we further decrease $D$. \label{fig:LangevinTravel}}
\end{figure}
To quantitatively characterise the emergence of travelling-wave states, we introduce the flux
\begin{equation}
    J_x =\left \langle \rho_A(x) \frac{\partial \rho_B(x)}{\partial x} - \rho_B(x) \frac{\partial \rho_A(x)}{\partial x} \right\rangle,
\end{equation}
which serves as an effective order parameter for self-propelled patterns. The average is taken over both time and space. Physically, this quantity measures the relative phase shift between the density profiles of the two species and captures the presence of a net directional transport. For standing-wave configurations, where the densities oscillate in time but remain spatially symmetric, the contributions to $J_x$ cancel, yielding a vanishing flux. In contrast, for travelling-wave states, a persistent phase difference between $\rho_A$ and $\rho_B$ leads to a nonzero value of $J_x$, signalling the emergence of polar, self-propelled patterns. The dependence of the flux magnitude $|J_x|$ on the diffusion coefficient $D$ is shown in Fig.~\ref{fig:LangevinTravel}. As $D$ is decreased below the critical value $D_{cH}$, $|J_x|$ increases continuously from zero, indicating the smooth onset of directed motion. This behaviour is consistent with the supercritical Hopf bifurcation of the $q=2$ mode, in which a stable limit cycle emerges continuously from the homogeneous state. Thus, the growth of $|J_x|$ provides a direct quantitative signature of the transition from stationary or standing-wave dynamics to travelling-wave behaviour. If we reduce $D$ further, a SNIC-like bifurcation takes place as is evident in Fig.~\ref{fig:LangevinTravel}.
\section{Conclusion}
In this work, we have investigated a nonreciprocal extension of the MVE, combining analytical methods with numerical simulations to uncover the role of asymmetric interactions in shaping collective dynamics. Our analysis demonstrates that nonreciprocity fundamentally alters the stability and dynamical behaviour of mean-field systems, with the nature of these effects depending sensitively on both the strength and spatial structure of the asymmetry.
For the case of constant nonreciprocity, we find that the primary effect is a shift in the critical diffusion coefficient governing the onset of instability. In this regime, the system exhibits only stationary instabilities, and the homogeneous state loses stability through the growth of non-oscillatory modes. This behaviour reflects the fact that weak, spatially uniform nonreciprocity modifies the effective interaction strength but does not introduce the phase lag necessary to generate time-dependent collective motion.

In contrast, when the nonreciprocity is spatially modulated, the system exhibits a much richer range of dynamical phenomena. Our LSA reveals that the symmetry of the nonreciprocity function plays a crucial role: the even component of the spatial modulation breaks action-reaction symmetry and can generate Hopf instabilities, whereas purely odd modulations provide reciprocal control cases and yield only stationary instabilities here. These predictions are supported by pseudo-spectral simulations of the continuum equations, which show the emergence of both standing and travelling wave states. Beyond the linear regime, our numerical analysis uncovers distinct types of bifurcation behaviour. In particular, we identify subcritical Hopf bifurcations associated with discontinuous transitions, as well as supercritical Hopf bifurcations. The Landau coefficients together determine the nature of the oscillatory state. For travelling waves, the sign of the real part of the self-saturation coefficient $g_s$ controls the bifurcation: $\text{Re}[g_s]<0$ corresponds to a subcritical Hopf bifurcation, while $\text{Re}[g_s]>0$ indicates a supercritical one. For standing waves, the relevant criterion involves the combined effect of self- and cross-saturation, with $\text{Re}[g_s + g_c]< 0$ signalling a subcritical Hopf bifurcation and $\text{Re}[g_s + g_c] > 0$ corresponding to a supercritical bifurcation. 
Notably, we demonstrate that travelling waves can emerge even in the weak-nonreciprocity regime, without requiring explicit run-and-chase mechanisms typically invoked in active matter systems. This highlights a minimal route to self-organized directed motion driven purely by spatially structured asymmetry.

To validate the continuum description, we perform Langevin dynamics simulations of the underlying stochastic particle system and find excellent agreement with the PDE results across all regimes considered. In particular, the persistence of travelling-wave states at the particle level confirms that these phenomena are not artefacts of the mean-field approximation. We further characterise the onset of travelling waves using a flux-based order parameter, which captures the transition from symmetric oscillations to directed motion and provides a clear signature of the underlying Hopf bifurcation.

Several directions naturally follow from this work. A first important step would be to derive the nonlinear Landau coefficients systematically from the microscopic interaction kernels, thereby enabling a complete phase diagram of nonreciprocal McKean-Vlasov systems in terms of the potential, spatial modulation of asymmetry, and diffusion strength. It would also be valuable to move beyond the weak-nonreciprocity regime considered here, where stronger asymmetries may generate additional instabilities, multistability, hysteresis, or new routes to directed motion. Extending the analysis to higher-dimensional domains is another promising direction, since fluctuations, defects, and geometry can qualitatively alter the stability of standing and travelling wave states. Finally, finite-size effects and noise-induced transitions in the underlying particle system deserve closer study, particularly near subcritical Hopf and SNIC-like bifurcations, where rare events may control switching between homogeneous, stationary, and oscillatory phases. Together, these extensions would deepen the connection between nonreciprocal mean-field theory and experimentally relevant systems in active matter, synchronization, and social dynamics.
\section{Acknowledgments}
The authors acknowledge financial support from the Department of Atomic Energy, India through the Project (RIN4001-SPS).

\bibliography{references}

\end{document}